# State Analysis and Aggregation Study for Multicast-based Micro Mobility

Ahmed Helmy
Electrical Engineering – Systems Department
University of Southern California, Los Angeles, CA 90089
helmy@ceng.usc.edu

## ABSTRACT

IP mobility addresses the problem of changing the network point-of-attachment transparently during movement. Mobile IP is the proposed standard by IETF. Several studies, however, have shown that Mobile IP has several drawbacks, such as triangle routing and poor handoff performance. Multicast-based mobility has been proposed as a promising solution to the above problems, incurring less end-to-end delays and fast smooth handoff. Nonetheless, such architecture suffers from multicast state scalability problems with the growth in number of mobile nodes. This architecture also requires ubiquitous multicast deployment and more complex security measures.

To alleviate these problems, we propose an intra-domain multicast-based mobility solution. A mobility proxy allocates a multicast address for each mobile that moves to its domain. The mobile uses this multicast address within a domain for micro mobility. Also, aggregation is considered to reduce the multicast state. We conduct multicast state analysis to study the efficiency of several aggregation techniques. We use extensive simulation to evaluate our protocol's performance over a variety of real and generated topologies. We take aggregation gain as metric for our evaluation.

Our simulation results show that in general leaky aggregation obtains better gains than perfect aggregation. Also, we notice that aggregation gain increases with the increase in number of visiting mobile nodes and with the decrease in number of mobility proxies within a domain.

## 1. INTRODUCTION

IP mobility addresses the problem of changing the network point-of-attachment transparently during movement. Mobile IP[4][5] is the proposed standard by IETF. However, several studies [1][3][7] have shown that Mobile IP has several drawbacks ranging from triangle routing and its effect on network overhead and end-to-end delays, to poor performance during handoff due to communication overhead with the home agent.

Multicast-based mobility [1][2] has been proposed as a promising solution to the above problems, incurring less end-to-end delays and fast smooth handoff. In such architecture, the mobile node is assigned a multicast address to which it joins through base stations it visits throughout its movement. Handoff is performed through standard multicast join/prune mechanisms.

In a previous route-analysis simulation study [1] we have shown that, on average, multicast-based mobility experiences around half the network overhead, end-to-end delay and handoff delays experienced by basic Mobile IP, and clearly outperforms its variants even with optimizations.

Nonetheless, the multicast-based architecture for inter-domain mobility suffers from several problems concerning multicast state scalability with the growth in number of mobile nodes, and, subsequently, number of groups. The architecture also requires ubiquitous multicast deployment and more complex security measures.

To alleviate these problems, we propose an intra-domain multicast-based mobility solution. In this architecture, a mobile node is assigned a multicast address within a domain that it uses for micro mobility. The allocated multicast address is locally scoped (i.e., unique only domain-wide). This allows for a domain-wide address allocation scheme, in which a mobility proxy (or a group of proxies) allocates multicast addresses for visiting mobiles. These addresses are locally-scoped and are used temporarily by the mobiles for micro mobility while moving within the domain. The mobile proxy performs inter-domain mobility on behalf of the visiting mobile, then tunnels the packets *multicast* to the mobile. The multicast address of a mobile does not change throughout its movement within the domain.

Since the multicast addresses are locally-scoped and the joins go through the mobility proxy, the multicast address allocation scheme is performed per-domain (as opposed to requiring an inter-domain architecture). Also, this provides potential for multicast state aggregation opportunities. In addition, we conduct multicast state analysis to quantify aggregation gains over various topologies for random movement patterns. We investigate two kinds of bit-wise aggregation; perfect and leaky aggregation. Our results show that leaky aggregation leads to better aggregation gain at the expense of extra unnecessary traffic. Also, we note that aggregation gain reduces with the increase in number of mobility proxies, increase in number of nodes in the network and increase in the average node degree.

The rest of this paper is organized as follows. Section 2 presents related work. Section 3 provides an overview of multicast-based mobility, while Section 4 discusses the issues associated with inter-domain multicast-based mobility and motivates our approach for multicast-based micro mobility. Section 5 presents our basic architecture and Section 6 provides architectural discussion. State aggregation is explained in Section 7. Simulation results and their analysis are presented in Section 8. Section 9 concludes and presents future work.

## 2. RELATED WORK

Several architectures have been proposed to provide IP mobility support. Work by the IETF on Mobile IP (MIP) is given in [4]. In MIP a mobile node (MN) is assigned a permanent *home* address and a *home agent* (HA) in its home subnet. When the MN moves to another *foreign* subnet, it acquires a temporary care-of-address (COA) through a foreign agent (FA). The MN informs the HA of its COA through a *registration* process. From that point on, packets destined to the MN's home address are sent first to the home network and are *tunneled* to the MN. This is known as the *triangle routing* problem, which is the major drawback of the basic MIP. A proposed mechanism, known as *route optimization*, attempts to avoid triangle routing. In [6][5] route optimization is achieved by sending *binding updates*, containing the current COA of the MN to the correspondent node (CN). However, communication overhead during handoff is still high rendering MIP unsuitable for *micro* mobility and causing it to be inadequate for audio applications.

Hierarchical mobility was proposed in [7] that defines three hierarchical levels of mobility; local, administrative domain and global mobility. This scheme proposes to use MIP for the global mobility, while using *subnet foreign agents* and *domain foreign agents* for the other levels. It is not clear, however, how this hierarchy will be formed or how it adapts to network dynamics, partitions or router failures. In [8] an end-to-end architecture is proposed for IP mobility, based on dynamic DNS updates. Whenever the MN moves, it obtains a new IP-address and updates the DNS mapping for its host name. A migration process is required to maintain the connection. The transport protocol is aware of the mobility mode during the migration process. Such architecture avoids triangle routing. However, we believe that such architecture incurs handoff latency due to DNS update delays and migration delays. The end-to-end approach is geared toward TCP-based applications, but we may not suitable for real-time applications with stringent delay and jitter bounds.

The Daedalus project [3] proposes to tunnel the packets from the HA using a pre-arranged multicast group address. The base station, to which the MN is currently connected, and its neighboring base stations (BSs), join that group and get the data packets over the multicast tree. It is not clear how the scheme performs in larger wide-area topologies. This approach suffers from the triangle routing problem; packets are sent to the HA first and then to the MN.

An approach for providing mobility support using multicast (MSM-IP) is presented in [2]. In this approach, each MN is assigned a unique multicast address. Packets sent to the MN are destined to that multicast address and flow down the multicast distribution tree to the MN. This is similar, in concept, to the Daedalus project approach. However, it is the CN (not the HA) that tunnels the packets using the multicast address. This approach avoids triangle routing, in addition to reducing handoff latency and packet loss. A hierarchy of servers is proposed for location management. Such hierarchy is complex, susceptible to failures, and imposes restrictions of placement of the Rendezvous Point in PIM-SM as an underlying multicast protocol.

In [1] we propose another architecture for multicast-based mobility. However, in [1] we propose a start-up phase that is a minor modification to Mobile IP to implement our protocol. In addition, by using binding updates and using the destination option in IPv6, we avoid potential MSM-IP problems with TCP (and other protocols) due to the use of multicast addresses for the MN. The multicast address is used within the network for packet routing, but the applications are only aware of the permanent unicast home address of the MN. Hence, no change to the application or transport protocol is needed. Our analysis quantifies the advantage gained for multicast-based mobility.

All such inter-domain multicast-based approaches, however, suffer from several issues, including scalability of multicast state, address allocation and dependency on inter-domain multicast. We discuss these issues and address them in this paper.

In the area of micro-mobility there exists several related architectures. We discuss the most prominent here. In [18] the HAWAII architecture is proposed. This architecture requires changes to the unicast routing protocols (including every router) and adds complexity in creating new unicast routing entries. In [17] cellular IP is proposed for handling micro mobility within LANs (i.e., layer 2 subnets). The MN is assumed to move between switches or base stations within the same LAN. The LAN is assumed to be connected to a gateway (i.e., router) that floods periodic beacons. Cellular IP was geared towards paging services. This approach is not sufficient for a domain-wide all-IP mobility architecture.

Other related work lies in the area of multicast state aggregation. The work in [19] proposes to use per interface approach for aggregation. This approach, however, benefits from having a large number of members to the group, which does not apply in our case. In [20] the concept of leaky aggregation was thoroughly studied for wide-area multicast routing. Here, we do study leaky aggregation as applied to our architecture.

## 3. Multicast-based Mobility

In multicast-based mobility, each mobile node (MN) is assigned a multicast address. The MN, throughout its movement, would join this multicast address through the locations it visits. Nodes wishing to send to the MN send their packets to a *multicast* address, instead of sending their packets to a unicast address. Because the movement will be to a geographical vicinity, it is highly likely that the join from the new location (to which the mobile has recently moved) will traverse a small number of hops to reach the already-established multicast distribution tree. Hence, performance during handoff will be improved drastically. In order to send packets to the multicast address of the mobile node, the correspondent nodes need to obtain this multicast address. This is performed during the start-up phase, where the MN notifies the CN of its multicast using *binding update* similar to MIPv6 [5]. However, unlike MIPv6, the binding update occurs only once during the initial establishment of communication, not with every move.

An overview of this architecture is given in Figure 1. As the MN moves, it joins to the assigned multicast address through the new base station. Once the MN starts receiving packets through the new location, it sends a prune message to the old base station to stop the flow of the packets down that path. Thus completing the smooth handoff process.

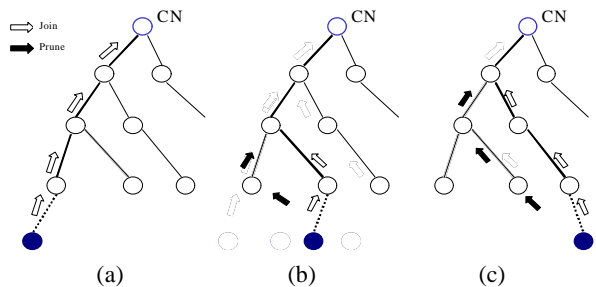

**Figure 1.** Multicast-based mobility. As the MN moves, as in (b) and (c), the MN joins the distribution tree through the new location and prunes through the old location.

In our earlier work [1] we have conducted extensive simulations to compare the performance of multicast-based mobility to Mobile IP and its variants. Here, we summarize some of our results from that study.

The protocols compared were Mobile IP (MIP) [4], Mobile IPv6 with route optimization [5], and a variant thereof where, during handoff, packets to the new base station are forwarded from the old base station to reduce handoff latency (we refer to this approach as the *previous location* approach). The performance metrics used for comparison were end-to-end delay, handoff delay and network overhead (which we omit for brevity).

End-to-end delays were measured to evaluate the effect of triangle routing. In MIP, data packets are routed to the HA then are tunneled to the FA; i.e., packets traverse the path *'A+B'* shown in Figure 2 (a). In our multicast-based approach, data packets take the shortest path from the CN to the MN; i.e., packets traverse path *'C'* in the figure. Hence, the ratio of the end-to-end paths in both cases is *'r=(A+B)/C'*. As for handoff performance, delay during handoff is a function of the path traversed by control messages to bring the data to the new base station. In MIP, registration request is sent to the HA; i.e., traverses path *'B'*, whereas in MIPv6 the binding updates are sent to the CN; i.e., path *'C'*. In the previous location approach, updates need to be sent to the previous base station; this is labeled *'P'* in Figure 2 (b), and for our multicast-based mobility approach join messages need to reach the multicast tree; this is labeled *'L'* in the figure.

We have conducted route-based analysis to evaluate the above metrics. Simulation was carried out for various topologies (partial table is given in Table 1), and for various movement models (here we only consider the *cluster* movement, which is, we believe, more representative of the average[1]).

Results of our simulation are shown in Figure 3 and Figure 4. The average end-to-end delay ratio *'r'* was found to be '2.22', whereas the average handoff delay ratio, with respect to our multicast-based mobility (M&M) approach, was: 1.44 for previous location (PL), 2.3 for MIP and 2.43 for MIPv6 as shown in Figure 4.

In sum, our analysis showed that our multicast-based mobility approach achieves notable improvement in both the end-to-end delay and handoff performance. Our work was the first to measure such improvements quantitatively over large scale networks.

In spite of such promise, we believe that many issues need to be addressed to realize multicast-based mobility in today's Internet. Such issues, along with our proposed solution, are discussed next.

| name | nodes | links | av deg | name | nodes | links | av deg |
|---|---|---|---|---|---|---|---|
| ARPA | 47 | 68 | 2.89 | 1008.1 | 1008 | 1399 | 2.78 |
| ts50 | 50 | 89 | 3.63 | 1008.2 | 1008 | 2581 | 5.12 |
| ts100 | 100 | 185 | 3.7 | 1008.3 | 1008 | 3787 | 7.51 |
| ts150 | 150 | 276 | 3.71 | ti1000 | 1000 | 1405 | 2.81 |
| ts200 | 200 | 372 | 3.72 | ti5000 | 5000 | 7084 | 2.83 |
| ts250 | 250 | 463 | 3.72 | mbone | 3927 | 7555 | 3.85 |
| ts300 | 300 | 559 | 3.73 | Mbone | 4179 | 8549 | 4.09 |
| ts1000 | 1000 | 1819 | 3.64 | AS | 4830 | 9077 | 3.76 |

Table 1. Simulated topologies (ts: transit-stub)

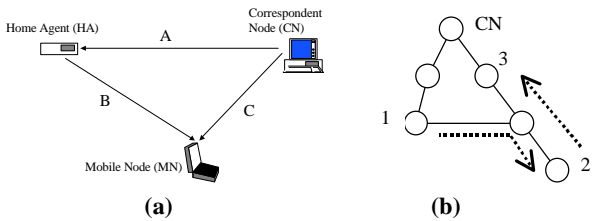

**Figure 2. (a)** Ratio *'r'*= *(A+B)/C.* **(b)** As the MN moves from node 1 to 2, added links '*L*' is 3 and links to previous location '*P*' (dashed lines) is 2. As it moves from 2 to 3, ***L****=0, P=2*.

---

[1] In the cluster movement the MN is allowed to connect randomly to only one of 6 nodes that are likely to fall within the same cluster as the MN.

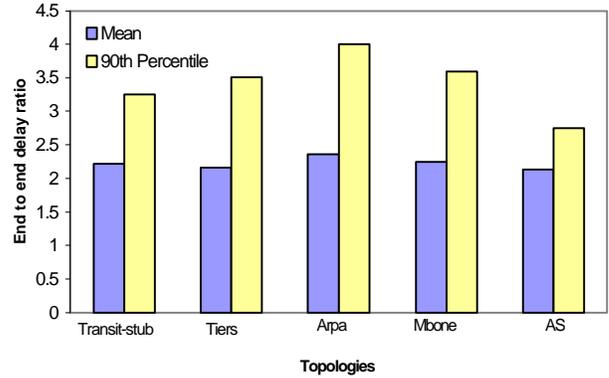

**Figure 3.** End to end ratio for the cluster movement

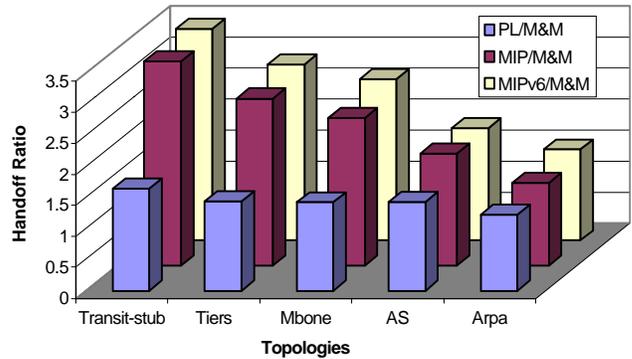

**Figure 4.** Average handoff delay ratios (PL: previous location approach, M&M: multicast-based mobility).

## 4. Issues

Several issues are involved in the deployment of the multicast-based mobility architecture over wide-area networks (i.e., for inter-domain mobility). In this section we focus on what we think are the major concerns: scalability of multicast state, multicast address allocation, requiring ubiquitous deployment of multicast, and security overhead during handoff. We discuss these issues and present an architecture offering a common solution to alleviate and hopefully eliminate some of these issues.

### 4.1 Scalability of Multicast State

Each mobile node is assigned a multicast address to which it joins throughout its movement. The state created in the routers en-route from the MN to the CN is source-group (*S,G*) specific state.

With the growth in number of mobile nodes, and subsequently, number of groups (*G*), the number of states kept in the router increases. In general, if there are '*x*' MNs, each communicating with '*y*' CNs on average, then routers in the network should create '*x.y*' (*S,G*) states. This obviously does not scale for inter-domain mobility.

### 4.2 Multicast Address Allocation

The problem of multicast address allocation is a research problem in the Internet community [10]. This problem will be exasperated by requiring each MN to be assigned a globally unique multicast address.

Aside from the fact that the multicast address space is restricted for IPv4, using a global multicast address for each MN may be wasteful and requiring uniqueness may not be practical. There is not current scheme that would ensure such allocation.

### 4.3 Ubiquitous Multicast Deployment

In order to implement inter-domain multicast-based mobility, inter-domain multicast routing needs to be in place. Unfortunately, this requirement restricts the applicability of our inter-domain mobility architecture, especially in the absence of an interoperability interim architecture.

### 4.4 Security Overhead

Security is critical for mobility support, where the continuous movement and change of attachment point is part of the normal operation. Such setting is prone to *remote redirection* attacks, where a malicious node redirects to itself packets that were originally destined to the mobile node. In general, authentication should be used with any message revealing information about the mobile node. The problem is even more complex with multicast, where any node may join the multicast address as per the IP-multicast host model.

These security measures are complex and may incur a lot of overhead. If such measures are invoked with every handoff, however, it may overshadow the benefits of efficient handoff mechanisms, including multicast-based handoff.

To alleviate these problems, we propose an intra-domain multicast-based mobility solution, presented in the following section, along with a discussion of how the proposed architecture addresses the above issues.

### 5. Intra-domain Architectural Overview

Similar to the concept of multicast-based mobility for the inter-domain case, in this (intra-domain) architecture, a mobile node is assigned a multicast address to which it joins while moving. However, the multicast address is assigned only within a domain (e.g., autonomous system or AS) and is used for intra-domain micro mobility. While moving between domains, an inter-domain mobility protocol is invoked (e.g., Mobile IP). We do not assume a specific protocol for inter-domain, only that such a protocol exists. For the sake of illustration, we take MIP as an example for inter-domain mobility protocol, when needed.

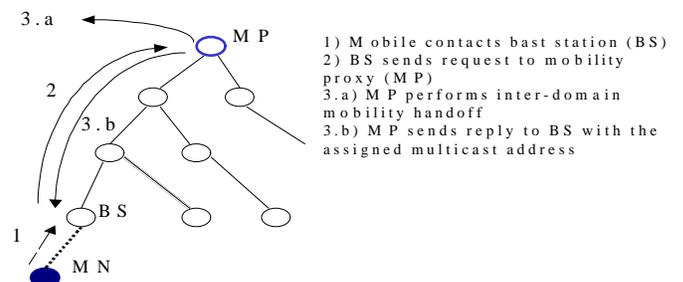

**Figure 5.** Sequence of actions as the mobile node moves into a domain.

When a mobile node moves into a new domain, it contacts the entry point base station (the first base station it encounters). This entry point base station

(BS) performs the necessary per-domain authentication and security measures, then assigns a unicast care-of-address (CoA) for the mobile node to use in that subnet. As shown in Figure 5, the BS then sends a *request* message to the mobility proxy (MP) to obtain a multicast address for the visiting MN[2]. The request message includes the home address of the mobile node and its home agent's address. Upon receiving the request the MP performs two tasks. The first is to execute the inter-domain handoff on behalf of the MN. In the case of Mobile IP, for example, this means that the MP registers its own address (as the new CoA) with the MN's home agent. The second task is for the MP to assign a multicast address for the visiting MN, send a *reply* message to the BS and keep record of this mapping. The mapping is used for packet encapsulation later on.

Once this step is complete, the visiting MN joins the assigned multicast address ($G$). The joins are sent to ($MP,G$) and are processed as per the underlying multicast routing[3]. The MN continues to move within the same domain using the same multicast address. The assigned multicast address is locally scoped to the domain. Handoff is performed using standard join/prune mechanisms and only lightweight intra-domain security is required in this case.

When packets are sent to the MN, they are forwarded to the MP using inter-domain mobility. The packets are then encapsulated by the MP, based on the mapping, and multicast to the MN. For example, in Mobile IP the home agent encapsulates the packets and sends them to the MP. The MP looks into the inner header to know the home address of the destination, performs the mapping, strips off the outer header and encapsulates the inner packet with multicast header. The packets flow from the MP down the multicast tree to the MN.

## 6. Architectural Discussion

We would like to point out several issues with the above architecture.

**Scalability** Our architecture attempts to address the limitations of the inter-domain multicast-based mobility. In terms of multicast state scalability we note that the multicast state growth is $O(G)$ for the architecture presented in this study, as opposed to $O(SxG)$ in [1][2]. However, there is still some concern for state concentration on certain paths (i.e., in certain routers) in the network. To further improve scalability of multicast state we investigate several aggregation techniques in the next section. We believe this is worthy of study as the problem of state scalability is an important one.

**Address allocation** Performed by the mobility proxies on a per-domain basis, the multicast address assignment is now a local mechanism, and the multicast addresses are locally scoped within the domain. This facilitates address allocation, in addition to providing per-domain privacy as the multicast packets are not forwarded out of the domain.

**Incremental multicast deployment** Based on per-domain approach, our architecture allows for incremental deployment of multicast. This way, the best handoff performance can be attained using our architecture[4] without requiring inter-domain multicast.

**Security overhead** Lightweight intra-domain security mechanisms may be used while moving

---

[2] The MP is typically placed at or near the domain's border router, but it may be placed anywhere in the domain.
[3] Note that this is not a source-group state. Rather, it is for all sources sending to the MN ($G$). This is similar in concept to the (*,$G$) tree established towards the Rendezvous Point (RP) in PIM-SM [9], but can be achieved using any multicast routing protocol.
[4] This has been shown through extensive simulations in[1].

within a domain, thus reducing security overhead during handoff.

**Robustness** To avoid single-point-of-failure scenarios (especially for the mobility proxy) we provide several mechanisms to enhance our protocol robustness. Instead of having only one mobility proxy (MP) per-domain, we propose to have multiple MPs (typically, five to ten per-domain). These MPs are typically placed/configured at the border of the domain or at the center of the network[5]. On average, this achieves reasonably performance[6].

Each MP sends periodic *liveness* messages to a well-known domain-specific group called *MP-announcement-group*. All base station routers join this group and receive the liveness messages. Each such router maintains a *live-MP* list and maintains a timer for each MP that is reset by the liveness message from that MP. When a base station router is first contacted by a visiting MN, it performs a hash procedure to select one of the MPs from the MP-list. We use a hash procedure to avoid distributing explicit mapping, which does not scale well. The hash procedure assigns a weight to each MP$i$ using hash(MNaddress, MP$i$). Then selects the highest weight MP to which it sends the request message. This scheme has two advantages. First, it distributes the visiting MNs equally over the MP-list. Second, if a MP fails only those MNs that hashed to it are re-hashed, other MNs are not affected. See [21] for more detail on such algorithm. Moreover, if a new MP is added to the pool of MPs (i.e., the change in the list was not caused by failure) no re-hashing is done.

Failure of a MP is detected by the base station routers when the MP timer expires. If the router uses the failed MP for some of its MNs, it does the re-hashing for those MNs to select a live MP[7].

## 7. State Aggregation

One of the main problems with multicast-based mobility is scalability of multicast state with the increase in number of visiting mobile nodes. This is especially a problem where state concentration is expected to occur, as in the mobility proxies. Hence, we propose to use multicast state aggregation to reduce the state requirement in the network routers. There are several techniques for state aggregation. Aggregation used in unicast routing for the Internet is *prefix* aggregation. That is, two states can be aggregated if they have the same unicast address prefix. This is very efficient for aggregating domain routing information since a domain/subnet/LAN has a specific unicast prefix. It is not clear if this benefit will also apply for the multicast case, since multicast addresses are not geographically significant. Another kind of aggregation is the bit-wise aggregation.

Intuitively, bitwise aggregation provides more opportunity for aggregation, hence we expect it, on average, to provide better aggregation. However, a deeper look at the two schemes shows us scenarios where prefix aggregation leads to more aggregation than bitwise. For example, a sequence of {0,4,1,2,3} leads to 3 states in the bitwise aggregation, whereas with prefix aggregation it leads to 2 states. We perform simple analysis to understand the behavior of the two schemes.

---

[5] The center(s) of the network are the nodes with min(max) distance to reach any node in the network. This can be identified by the network admin at the point of configuration.

[6] For traffic originating outside the domain, the packets go through the border router anyway. As for the network center, studies [16] have shown that it performs well.

[7] The MN keeps the multicast and MP addresses. In case of MP failure during handoff, the new BS gets the multicast and MP addresses from the MN, and checks if the MP is alive. If it is not, then the BS performs the hashing and obtains a new live MP to which it sends a request, so on. This mechanism obviates the need for mapping state replication among MPs, which would require a lot of network overhead and protocol complexity. If the MN crashes during handoff (we assume it is hard configured with its home address at least), then the new BS, performs the hashing, gets the MP address and sends a request to the MP.

We define the aggregation ratio as the ratio of number of states before aggregation to the number of states after aggregation. Figure 6 shows the aggregation ratio for in-order numbers. Both aggregation techniques have identical behavior.

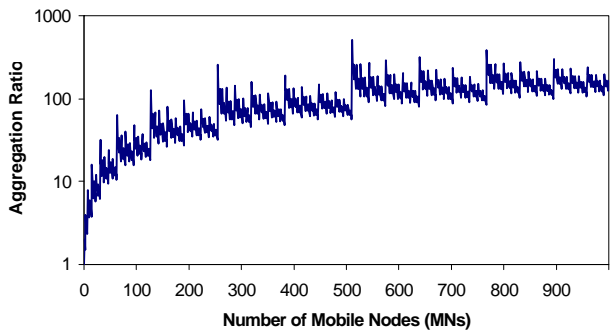

**Figure 6.** Aggregation ratio for in-sequence numbers. Identical gain for bitwise and prefix aggregation.

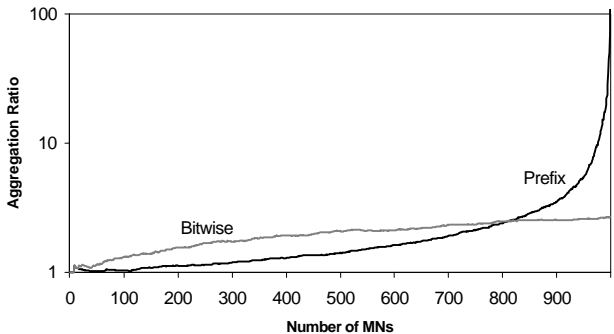

**Figure 7.** Aggregation ratio for random numbers. Bitwise aggregation outperforms prefix aggregation until 80% of the total number population is reached.

Figure 7 shows the aggregation ratio when the numbers are random. That is, out of 0 to 999, distinct numbers are chosen randomly until the whole number population is covered[8]. The random arrival of addresses is a more likely scenario, since mobile nodes arrive at different entry points and experience various movement patterns. The following table presents the results:

|  | Av. prefix | Av. bitwise | Av. bitwise/prefix |
|---|---|---|---|
| 80% population | 1.40 | 1.84 | 1.32 |
| 100% population | 2.48 | 1.98 | 1.19 |

It is interesting to note the cross over point at 80% population. The overall average aggregation ratio for prefix aggregation is '2.48' and for bitwise is '1.98'. However, up to 80% of the population bitwise aggregation outperforms prefix aggregation by a factor of 1.32. Hence, we use bitwise aggregation for further analysis.

Another classification of multicast aggregation is perfect vs. leaky aggregation. Since the multicast state consists of {source, group, incoming interface, outgoing interface list}, all fields must be compared in order to aggregate groups. The source of the multicast is the MP and the incoming interface is that pointing towards the MP. As for the outgoing interface list (in our case there's only one outgoing interface), if the interfaces are the same, then this is called perfect aggregation. If the states are aggregated even though the interfaces may be different, then this is called leaky aggregation, and it achieves better aggregation at the expense of extra network overhead. The data packets in this case may be sent over an extra link that does not reach a receiver.

We investigate both leaky and perfect aggregation in our simulations.

## 8. Simulation and Analysis

The first step to attempt to solve the scalability problem of multicast state is to understand the distribution of these states in the routers. Then after we apply aggregation techniques to these states it is easier to point out the benefits of such techniques. Aggregation gain, in general, depends on several factors, including topology, MP placement, movement patterns, number of MNs, among others. Here, we attempt to investigate bitwise leaky and perfect aggregations over several topologies, with random movement patterns[9] and heuristic placement of MPs (at or near the border routers).

---
[8] We have also obtained similar results with several other simulation runs with random numbers.

**8.2 Simulation Setup**

We have used the network simulator (NS-2) [15]. Two sets of simulation scenarios were investigated. In the first set, 1000 MNs enter the domain at different random times, and move to random nodes within the domain at random times, each time joining through the new location and pruning through the old location. Any existing MN may move random number of movements to random locations before the next MN enters into a random entry point, thus capturing the dynamics of the tree. Up to 250k moves were simulated.

Another set of scenarios capture snapshot of the network. MNs enter the domain at random entry nodes and at random times, however they do not move, thus simulating a snapshot of the domain where nodes may exist at random locations. This approach allows us to scale our simulations to up to 250k MNs.

In both simulation scenarios we use different number of mobility proxies, ranging from 1 to 4 proxies (placed somewhat randomly, but mainly in well-connected nodes, that are likely to be border routers,

---

[9] Random movement establishes, on average, a lower bound on aggregation, since the state will be dispersed over the network. Other kinds of movement, such as neighbor and cluster movements [1], should exhibit better performance, especially if we limit the entry base stations

for example). Out of those proxies, with every move or new entry, the MN establishes new connections to a random number of proxies, and maintains a random number of already-existing connections (if any).

In each of the simulations the number of multicast states at each router in the network is recorded with every new entry or move.

We have simulated several topologies from Table 1 with nodes ranging from 47 nodes up to 1000 nodes, that are likely to represent intra-domain networks. We focus our results discussion on the general trends we have observed.

To sum, we aim to enhance our understanding of the multicast state distribution in the nodes, and the effectiveness of bitwise leaky and perfect aggregation techniques. The problem is studied across different dimensions; network size and number of mobility proxies. The metrics used are state distribution in nodes and aggregation ratio.

**8.2 Analysis and Results**

We first discuss analysis of a single topology with 100 nodes. This will act to illustrate our analysis method, and should enhance our understanding of multicast state distribution and aggregation gains. Then we present results for simulations over topologies with various number of nodes and various number of mobility proxies.

**- 100 Nodes with 1 MP:**

The first topology used for the simulation is that given in Figure 8, with 100 nodes and transit-stub structure[10]. One mobility proxy, placed at node 0, was used in the simulation.

---

to a subset of the network nodes, which we did not in our simulations.

[10] This topology was used in a previous studies[1] [22].

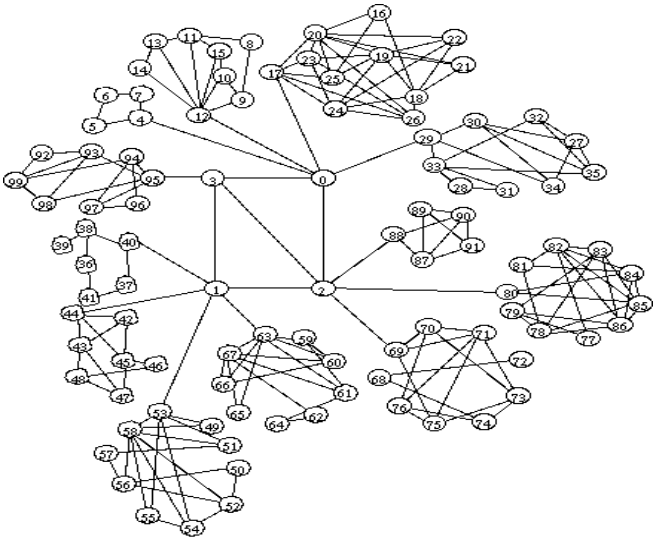

**Figure 8.** 100 node transit-stub topology (ts100).

The first scenario we discuss is for 1000 mobile nodes that enter the above topology at random points and move randomly to other nodes. We show simulation results for 40k moves. Figure 9 shows the multicast state distribution across the nodes (we only show the first 50 nodes for clarity. The graph starts at 250MNs for clarity). We notice that much of the multicast state in the network is concentrated at nodes 0,1,2 and 3 (i.e., the backbone nodes) as was expected.

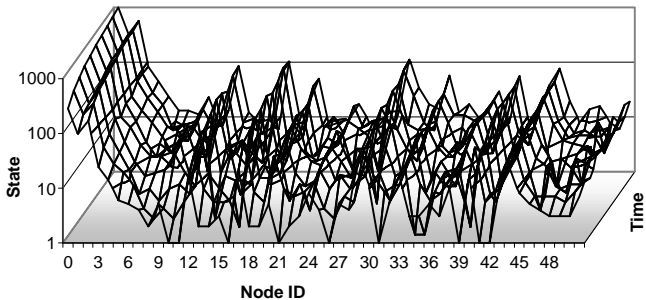

**Figure 9.** State distribution without aggregation

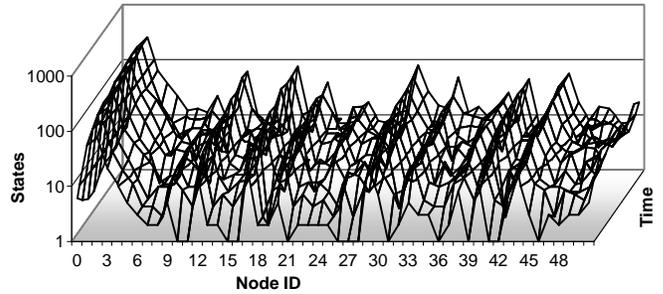

**Figure 10.** State distribution with leaky aggregation.

A similar simulation experiment was conducted and leaky aggregation was used. The state distribution across the nodes is shown in Figure 10. It is clear from the two previous graphs that the nodes where aggregation was most effective are those nodes with maximum state (nodes 0,1 and 2), whereas for the rest of the nodes the change was not that significant. For node 0, the MP, the aggregation ratio was so notable because the visiting MNs were assigned addresses in order. This brings up a question for the multiple MP case, which we will study later in this section.

We take a closer look at nodes 0,1 and 2 in Figure 11. We see that the number of states at node 0 dropped from above 250 to 1000 to below 10 states. Notable reduction in state was also observed for nodes 1 and 2. The overall number of states (90[th] percentile and average) over the 100 nodes is given in **Figure 12**. As shown, there is good improvement for the leaky aggregation case over the original case (factor of about 2 for average number of states and around 1.5 for 90[th] percentile). Note that the average number of states in the original case at a given time (in case of random movement) can be obtained from the simple equation: *No. MNs * Av.PathLen / No. Nodes,* where *No.MNs* is the number of mobile nodes (total of 1000 in our case), and *Av.PathLen* is the average path length in the topology (4 in our case), and *No.Nodes* is the number of nodes in the topology

(100 in this case). For example, the average number of states for 1000 MNs is 40. Also, we noticed a significant decrease in the *variance* of states in the nodes; i.e., aggregation leads a more balanced network in terms of state[11].

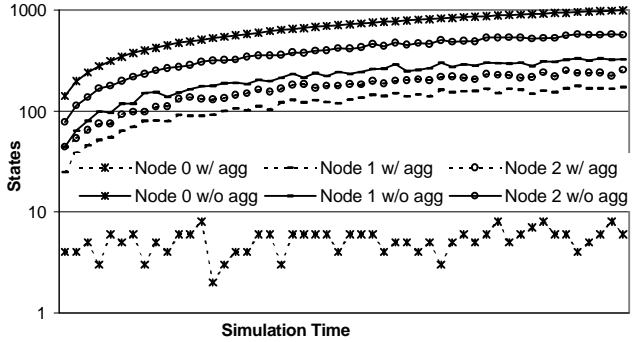

**Figure 11.** Number of states in nodes 0,1 and 2 (w/o agg: without aggregation, w/ agg: with aggregation)

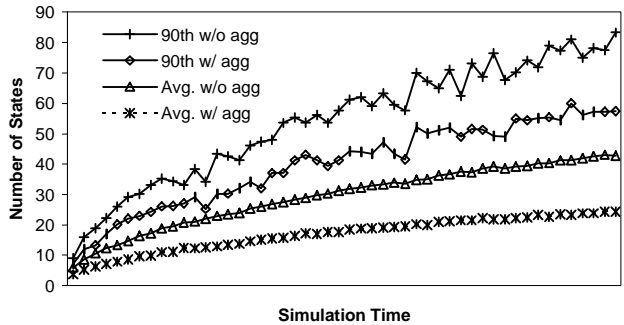

**Figure 12.** Overall average and 90[th] percentile.

The second scenario simulated was the snapshot scenario, for 250k MNs connecting at random points. The state distribution across time is given in Figure 13 (the data shown starts from 10k MNs for 50 nodes, for clarity). Again, we see concentration of the state at nodes 0 through 3, but also we observe surges in some other nodes (observe the darker areas of the graph).

So, we take a closer look at the state distribution at the end of simulation as given in Figure 14. This figure shows the multicast state distribution after 250k MNs (last snapshot). The average state per node is 10,830 states[12]. However, only 20% of the nodes had 10k or more states, and around 60% of the nodes have around 2500 states (i.e., 1% of the total number of MNs). This is a strong indication that the state distribution across nodes is not uniform (in fact it is quite skewed). Hence, there is potential to achieve good aggregation ratios in nodes with many states.

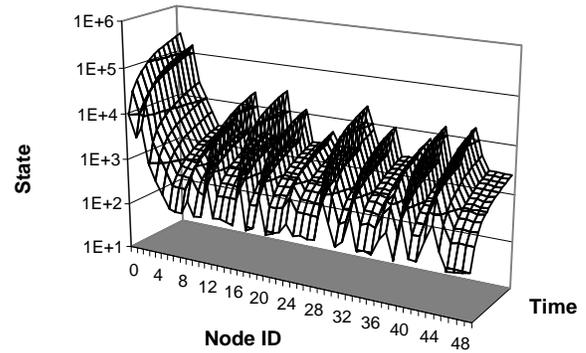

**Figure 13.** Distribution of multicast state across nodes and time, for 250k MNs (starting from 10k, only the first 50 nodes are shown for clarity.

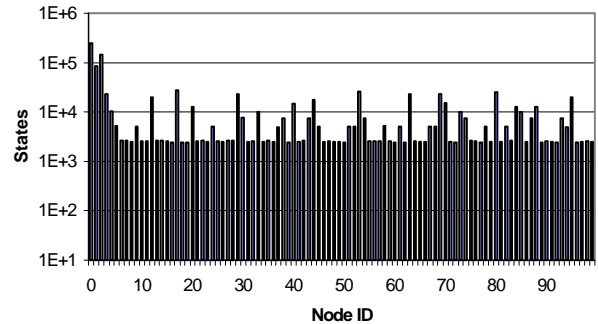

**Figure 14.** Number of multicast states indexed by the node ID after 250k MNs.

To further understand the aggregation performance, we apply both leaky and perfect aggregation techniques to the above scenario (up to 40k MNs). For both techniques, we measure the average aggregation ratio, as well as 90th percentile and maximum state ratios[13]. It is apparent that these aggregation ratios increase with the increase of number of MNs. Also, it is clear that the leaky aggregation achieves better aggregation ratios than perfect aggregation. More specifically, by the end of simulation, for leaky aggregation the average aggregation ratio was 1.88 (approaching 2 for large number of MNs), the 90th percentile ratio was 1.57, and the maximum state ratio is 4.37. Whereas for perfect aggregation these ratios were 1.37 (approaching 1.4), 1.23 and 2.17, respectively.

---

[11] In the case of more regular pattern of entry/movement (such as restricting the base station routers to 10% of all the nodes in the topology), we expect more concentration of states, and hence better aggregation ratio.

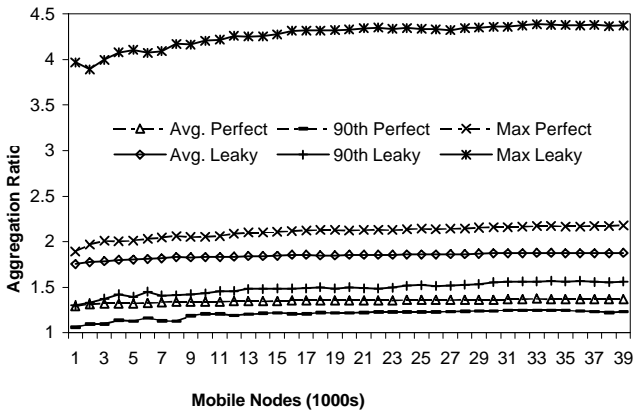

**Figure 15.** Aggregation ratios for leaky and perfect aggregation techniques.

**- Various Topologies with Multiple MPs:**

We now investigate both leaky and perfect aggregation techniques over several topologies (we only show topologies with 50 to 300 nodes for illustration. Similar trends were observed in other simulated topologies). We also explore and analyze aggregation trends with multiple mobility proxies (up to 4 proxies). The scenarios simulated are snapshot simulations, where 10k MNs connect at random nodes.

---

[12] Theoretically the average is 250k x 4 hops/100nodes = 10k states.
[13] Max state ratio=max state before agg/max state after agg, and similarly for the other ratios.

Aggregation ratio results for leaky aggregation are shown in Figure 16, and are summarized in the following table:

| MPs/Nodes | 50 | 100 | 150 | 200 | 250 | 300 |
|---|---|---|---|---|---|---|
| 1 | 1.99 | 1.84 | 1.71 | 1.71 | 1.64 | 1.63 |
| 2 | 1.48 | 1.40 | 1.36 | 1.36 | 1.36 | 1.32 |
| 3 | 1.38 | 1.33 | 1.31 | 1.31 | 1.29 | 1.28 |
| 4 | 1.33 | 1.29 | 1.27 | 1.27 | 1.26 | 1.25 |

As shown, the average aggregation ratio per node ranges from 1.25 (for the 300 node topology with 4 MPs) to 1.99 (for the 50 node topology with a single MP). The trend is clear; for the same number of visiting MNs, as the number of nodes in the topology increases, the state concentration in the nodes decreases and the aggregation ratio decreases. Furthermore, as the number of mobility proxies increase, the concentration of states in the nodes decrease and the aggregation ratio decreases.

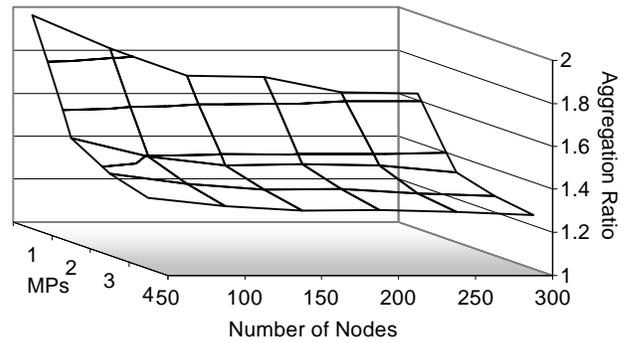

**Figure 16.** Aggregation ratio for leaky aggregation with various topologies and multiple MPs

Simulation results for the perfect aggregation are given in Figure 17, and are summarized in the following table:

| MPs/Nodes | 50 | 100 | 150 | 200 | 250 | 300 |
|---|---|---|---|---|---|---|
| 1 | 1.43 | 1.32 | 1.25 | 1.27 | 1.23 | 1.23 |
| 2 | 1.27 | 1.18 | 1.16 | 1.17 | 1.15 | 1.16 |
| 3 | 1.21 | 1.17 | 1.15 | 1.16 | 1.15 | 1.15 |
| 4 | 1.19 | 1.15 | 1.14 | 1.14 | 1.14 | 1.14 |

The average aggregation ratio ranges from 1.14 (for 300 node topology with 4 MPs) to 1.43 (for 50 node topology with a single MP). Evidently, leaky aggregation achieves better aggregation ratio. Also, the trends for both aggregation techniques are quite similar.

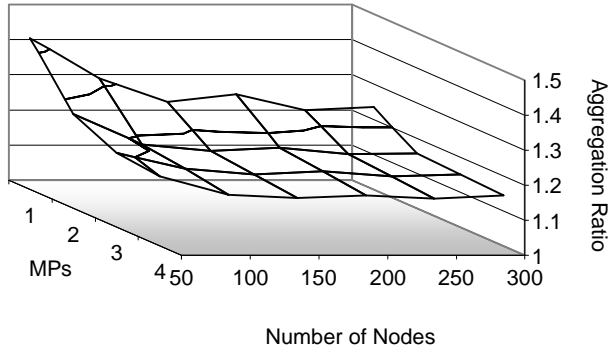

**Figure 17.** Aggregation ratio for perfect aggregation with various topologies and multiple MPs.

Other simulations conducted (not shown here for brevity) point out that the aggregation ratio increases as the average node degree decreases (i.e., the number of links decreases, hence the concentration of states increases).

## 9. CONCLUSION AND FUTURE WORK

In this paper, we presented a multicast-based protocol for supporting micro mobility. Our architecture is domain-based, in which mobility proxies assign visiting mobile nodes locally-scoped multicast addresses. A mobile node uses its assigned address during its movement throughout the domain. The handoff performance of such a scheme has been shown to outperform other IP mobility approaches. Our architecture addresses issues that have not been addressed before in previous multicast-based mobility approaches. Particularly, we addresses issues of multicast state scalability, multicast address allocation, incremental multicast deployment and overhead of security during handoff.

A thorough analysis of multicast state distribution among the nodes in the topology was presented based on extensive simulations. It was observed that, in general, multicast state tends to be distributed unevenly across the nodes in the topology. In addition, leaky and perfect bitwise aggregation techniques were studied and their aggregation ratio was evaluated for intra-domain topologies with varying number of mobility proxies.

Our findings indicate that the aggregation ratio increases with the increase in number of visiting mobile nodes. The maximum aggregation ratio is obtained for nodes with maximum multicast state. Also, aggregation helps balance the load of states kept at the nodes.

General trends were found for both aggregation techniques. As the number of mobility proxies increase, the multicast state in the routers tends to be more dispersed, and hence the aggregation ratio decreases. A similar observation is made with the increase in number of nodes in the topology. Also, as the average node degree in the topology decreases, the aggregation ratio increases.

Leaky aggregation clearly obtains better aggregation ratio than perfect aggregation, but at the expense of extra bandwidth. Leaky aggregation approaches an aggregation ratio of 2 whereas perfect aggregation approaches a ratio of 1.45 under similar conditions.

Several issues have not been addressed in this work that we plan to address in future work. Our current architecture requires the mobility proxies to maintain a mapping between the visiting MN address and the assigned multicast address. Can the scalability issues associated with this mapping be alleviated? Also, there is usually a trade off between state aggregation and forwarding performance. We have not studied such a tradeoff. Another issue is that of the join/prune protocol at the host side. IGMP was not optimized for wireless links that serve mobile nodes, especially with the MN being the only member of the group in that LAN. We shall investigate modifications to IGMP to render it more suitable to a wireless environment.

We also plan to conduct more detailed packet level simulations, including richer mobility patterns, deeper analysis of handoff performance and mechanistic protocol details. Especially, we plan to examine changes needed to the mobile node. Such changes should be minimized and should be power-efficient.

## REFERENCES

[1]  A. Helmy, "A Multicast-based Protocol for IP Mobility Support", ACM Second International Workshop on Networked Group Communication (NGC), Palo Alto, November 2000.

[2]  J. Mysore, V. Bharghavan, "A New Multicasting-based Architecture for Internet Host Mobility", *Proceedings of ACM MobiCom*, September 1997.

[3]  S. Seshan, H. Balakrishnan, R. Katz, "Handoffs in Cellular Wireless Networks: The Daedalus Implementation and Experience", *Kluwer Journal on Wireless Networks,* 1995.

[4]  C. Perkins, "IP Mobility Support", IETF *RFC 2002,* Oct 96.

[5]  C. Perkins and D. Johnson, "Mobility Support in IPv6", *Proceedings of MobiCom'96*, November 1996.

[6]  C. Perkins, D. Johnson, "Route Optimization in Mobile IP", *Internet Draft, IETF,* February 2000.

[7]  R. Caceres, V. Padmanabhan, "Fast and Scalable Handoffs for Wireless Internetworks" *MobiCom '96.*

[8]  A. Snoeren, H. Balakrishnan, "An End-to-End Approach to Host Mobility", *ACM MobiCom,* Aug 00.

[9]  D. Estrin, D. Farinacci, A. Helmy, D. Thaler, S. Deering, V. Jacobson, M. Handley, C. Liu, P. Sharma, "Protocol Independent Multicast – Sparse Mode: Protocol Specification", *IETF RFC 2362,* March '98.

[10] S. Kumar, P. Radoslavov, D. Thaler, C. Alaettinoglu, D. Estrin, M. Handley, "The MASC/BGMP Architecture for Inter-domain Multicast Routing", *Proceedings of ACM SIGCOMM,* August 1998.

[11] B. Cain, S. Deering, I. Kouvelas, A. Thyagarajan, "Internet Group Management Protocol, version3", *Internet draft of the IETF*, March 2000.

[12] K. Calvert, M. Doar, E. Zegura, "Modeling Internet Topology", *IEEE Comm,* p. 160-163, June 1997.

[13] G. Phillips, S. Shenker, H. Tangmunarunkit, "Scaling of Multicast Trees: Comments on the Chuang-Sirbu scaling law", *ACM SIGCOMM*, August 1999.

[14] P. Radoslavov, H. Tangmunarunkit, H. Yu, R. Govindan, S. Shenker, D. Estrin, "On characterizing network topologies and analyzing their impact on protocol design", *USC-CS-TR-00-731*, March 2000.

[15] L. Breslau, D. Estrin, K. Fall, S. Floyd, J. Heidemann, A. Helmy, P. Huang, S. McCanne, K. Varadhan, Y. Xu, H. Yu, "Advances in Network Simulation", *IEEE Computer,* vol. 33, No. 5, p. 59-67, May 2000.

[16] E. Fleury, Yih Huang, P.K. McKinley. "On the performance and feasibility of multicast core selection heuristics", 7th International Conference on Computer Communications and Networks, Page(s): 296 –303, 1998.

[17] A. Campbell, J. Gomez, S. Kim, A. Valko, C. Wan, Z. Turanyi, "Design, implementation, and evaluation of cellular IP" IEEE Personal Communications , Volume: 7 Issue: 4 , Page(s): 42 –49, Aug. 2000.

[18] R. Ramjee, T. La Porta, L. Salgarelli, S. Thuel, K. Varadhan, L. Li, "IP-based access network infrastructure for next-generation wireless data networks", IEEE Personal Communications , Volume: 7 Issue: 4 , Page(s): 34 –41, Aug. 2000.

[19] D. Thaler, M. Handley, "On the aggregatability of multicast forwarding state", IEEE INFOCOM 2000, Page(s): 1654 -1663 vol.3, 2000.

[20] P. Radoslavov, D. Estrin, R. Govindan, "Exploiting the Bandwidth-Memory Tradeoff in Multicast State Aggregation", USC-CS-TR99-697, Computer Science Department, USC, July 1999.

[21] D. Estrin, M. Handley, A. Helmy, P. Huang, D. Thaler, "A Dynamic Bootstrap Mechanism for Rendezvous-based Multicast Routing", Proceedings of IEEE INFOCOM '99, New York, March 1999.

[22] D. Zappala, "Alternate path routing for multicast", INFOCOM 2000, Page(s): 1576 -1585 vol.3, 2000.